\documentclass[]{aastex631}
\usepackage{amsmath}
\usepackage{pifont}

\newcommand{\eb}{\begin{equation}}
\newcommand{\ee}{\end{equation}}

\definecolor{rkka}{RGB}{219,66,32}
\definecolor{nsgreen}{rgb}{0.1,0.5,0.1}

\shorttitle{Distributions of wide stellar binary orbits}
\shortauthors{Makarov}

\begin{document}

\title{Distributions of wide binary stars in theory and in Gaia data:\\
I. Generalized Ambartsumian (1937) approach and the family of power-law distributions of eccentricity}

\correspondingauthor{Valeri V. Makarov}
\email{valeri.makarov@gmail.com}

\author[0000-0003-2336-7887]{Valeri V. Makarov}
\affiliation{U.S. Naval Observatory, 3450 Massachusetts Ave NW, Washington, DC 20392-5420, USA}

\begin{abstract}
The orbital parameter space of wide, weakly bound binary stars has been shaped by the still poorly known circumstances of their formation, as well as by subsequent dynamical evolution in parent clusters and in the field. The advance of the Gaia mission astrometry takes statistical studies of wide stellar systems to an unprecedented level of precision and scope. On the theoretical side of the problem, the old  approach proposed by Jeans and developed by Ambartsumian is revisited here. It is shown how certain simplifying assumptions about the phase density of binary systems in the framework of general copula distributions can lead to a family of analytical representations for the marginal distribution of orbital eccentricity, including accommodating and flexible power-law models. We further demonstrate the application of these models in forward Monte Carlo simulations of the measured motion angles between the relative velocity and separation vectors on the example of 170K listed Gaia binary systems, providing inference in the intrinsic distribution of orbital eccentricity.
 \end{abstract}

\section{Introduction} \label{int.sec}
The observed statistical distributions of directly measurable parameters of wide, resolved stellar binaries and hierarchical multiples, such as projected separation, position angle, and relative velocity,  are complex transformations of the underlying population distributions of basic physical parameters, including mass, semimajor axis, eccentricity, and the Euler angles of geometric orientation in a fixed inertial reference frame. The basic distributions of orbit size and eccentricity, and possible relations between them, are of special interest for cosmogony of such systems, finding evidence of long-term intrinsic evolution including the effects of tidal acceleration in the Galactic potential, interaction with the ambient stellar field, and, potentially, testing alternative concepts of gravitation in the weak field regime. 

Wide Jupiter-mass binary objects (JuMBOs) that has recently been discovered by the James Webb Space Telescope could be formed in dense  stellar clusters upon fortuitous encounters of regular planetary systems with alignment of the outer planets at the time of closest approach \citep{2024NatAs.tmp...78W}. Perhaps, similar result could be obtained from close encounters of soft stellar binaries, with the newly formed pair being wider than the initial intrinsic separations, and the eccentricity much higher than the typical values inherited from the core fragmentation mechanism in primordial proto-stellar disks. This wide-binary scenario predicts a marginal probability law of eccentricity steeply rising toward 1, i.e., the so-called super-thermal distributions, for the widest pairs. It has been assumed that binary stars are formed with a flat initial probability density of eccentricity, but dynamical interaction (encounters with other stars and binaries) within dense clusters leads to thermalisation of the distribution of energy and orbital momentum, and thus, the eccentricity. Recent studies have put this scenario to doubt \citep{2019ApJ...872..165G}. The limited information about eccentricities of closer and moderately separated binaries is more consistent with flatter distributions \citep{2013ARA&A..51..269D}. However, the domain of wider separations may be characterized with eccentricity rates rising toward 1, signifying a transition at approximately 1 KAU \citep{2017ApJS..230...15M}. Wider binaries within a cluster can emerge from dynamical capture and re-capture of previously disrupted systems \citep{2012ApJ...750...83P}, resulting in thermal or super-thermal distributions of eccentricity. These signatures, however, have not been confirmed with observational data for late-type binary systems \citep[e.g., ][]{2013ApJ...779...30G}.

\citet{1937AZh....14..207A} considered the previously proposed theoretical model by \citet{1919MNRAS..79..408J}, in which the general population of binaries is in the state of energy equipartition due to the previous dynamical interactions. In analogy with a thermally stable gas, the number density of systems in the six-dimensional phase space follows the exponential Boltzmann distribution law with a characteristic scaling temperature parameter, which is a function of only the energy. Under this assumption, the emerging probability density function (PDF) of eccentricity is independent of other orbital parameters, and is linear with eccentricity:
\eb 
{\rm PDF}[e]=2\,e.
\label{2e.eq}
\ee 
\citep{1919MNRAS..79..408J} also derived a theoretical distribution of orbital period $P$ (or orbital separation $a$), which turned out to be non-physical, i.e., indefinite-integral on the support $[0,\infty]$. The conclusion was that binary systems cannot follow Boltzmann's phase density distribution law. \citet{1937AZh....14..207A} redeemed this theoretical approach by noting that Boltzmann's law is not actually required to obtain a linear PDF$[e]$, which, by analogy with gas kinematics, is called the thermal distribution. Here we briefly recall this elegant and direct derivation, correcting some typos and stating unspecified assumptions in the original paper.

\section{Ambartsumian (1937)}
\label{amb.sec}
We begin with the space of six canonical Delaunay elements, which are defined as \citep{1999ssd..book.....M}:
\begin{eqnarray} \label{dela.eq}
    L & = & \hat\mu \sqrt{G(m_1+m_2)a} \nonumber \\
    \Gamma & = & L\,\sqrt{1-e^2} \nonumber \\
    H &=& \Gamma\,\cos\,i\nonumber \\
    l&=&M\nonumber \\
    \gamma&=&\omega \nonumber \\
    h&=&\Omega
\end{eqnarray}
where $\hat\mu=m_1\,m_2/(m_1+m_2)$ is the reduced mass, $m_1$ and $m_2$ are the masses of the primary and secondary components, $G$ is the constant of gravitation, $a$ is the semimajor axis, $e$ is the eccentricity, $i$ is the inclination angle to the chosen reference plane, $M$ is the mean anomaly, $\omega$ is the periastron argument angle in the orbital plane, and $\Omega$ is the longitude of the ascending node in the reference plane from the $X$ axis (vernal equinox in solar system dynamics). 

The more general and accommodating assumption considered by Ambartsumian is that the phase density of binary population in the 6D parameter phase is a function of only one element $L$, $D(L)$, and, therefore, it is independent of the other elements. The number of systems within an infinitesimal hyper-volume
of phase space is then
\eb 
N(L)=D(L)dL d\Gamma dH dl d\gamma dh.
\ee 
We note here in passing that Ambartsumian employed approximate relations for Delaunay parameters for the case $m_1\gg m_2$, which should not be used for binary systems with their distribution of mass ratios peaking above 0.5. The PDF of eccentricity is then readily derived from considering the number of systems with $\Gamma$ greater than a fixed value $\Gamma_0$:
\eb 
N(L|\Gamma>\Gamma_0)=\int D(L)dL\,\int_{\Gamma_0}^L d\Gamma\,\int_0^\Gamma dH\,\int dl\,\int d\gamma\,\int dh.
\label{D1.eq}
\ee 
We note here that the integration limits 0 and $\Gamma$ over $H$ imply that $\cos i$ varies between 0 and 1. This is not consistent with the traditional definition of $i$ as the inclination at the ascending node, which varies between 0 and $\pi$. This modification of inclination variable does not change the result of integration numerically apart from a constant factor of 2. Taking into account that the orbital orientation and phase angles are intrinsically random and uniformly distributed in the equatorial reference frame\footnote{The question of randomness of orbit orientation in the Galactic coordinate system has been investigated in the literature for many decades, with no significant evidence pointing otherwise} simplifies this integration to
\eb 
N(L|\Gamma>\Gamma_0)=16\pi^3\,\int D(L)dL\,\int_{\Gamma_0}^L \Gamma d\Gamma= 8\pi^3 \,\int D(L)(L^2-\Gamma_0^2)dL.
\ee 
Since $L^2-\Gamma_0^2=e_0^2$ from Eq. \ref{dela.eq}, and $N(L|\Gamma>\Gamma_0)=N(L|e<e_0)$,
we derive the cumulative distribution function (CDF) of eccentricity:
\eb 
{\rm CDF}[e]\propto e^2,
\label{therm.eq}
\ee 
which is equivalent to a PDF
\eb 
D_e(e)=2e.
\label{the.eq}
\ee 
This theoretical distribution can be tested with observations, which has been done in a number of publications \citep{2010ApJS..190....1R, 2017ApJS..230...15M, 2012AJ....144...54G, 2013ApJ...779...30G, 2020MNRAS.496..987T} in clusters and in the field. It implies that the rate of systems with $e>0.5$ should be three times higher than the rate of $e<0.5$ systems, and the sample mean eccentricity equals to 2/3, while the median eccentricity equals to $1/\sqrt{2}$.

\section{Generalization of Ambartsumian's approach} \label{met.sec}
Ambartsumian's approach can be further developed with a less restrictive and, possibly, more realistic assumption than the flat dependence of phase density on all orbital parameters except $L$. This assumption has been criticized in later literature, e.g., \citep{1975MNRAS.173..729H}. The explicit transition from the purely kinetic energy of gas molecules in thermal equilibrium to the total orbital energy of a population of binary stars is not obvious. A step further can be made if we assume that the population phase density is a function of both Delaunay parameters $L$ and $\Gamma$, thus accommodating an infinite set of possible distributions of eccentricity. Equation \ref{D.eq} can be re-written as
\eb 
N(L,\Gamma|\Gamma>\Gamma_0)=\int\int_{\Gamma_0}^L D(L,\Gamma)d\Gamma\,dL\, \int_0^\Gamma dH\,\int dl\,\int d\gamma\,\int dh.
\ee 
The two-dimensional distribution $D(L,\Gamma)$ can be modeled as copula distribution with a specific kernel function. According to Sklar's theorem\footnote{Weisstein, Eric W. "Sklar's Theorem." From MathWorld--A Wolfram Web Resource. \url{https://mathworld.wolfram.com/SklarsTheorem.html}}, any such distribution can be uniquely represented as a copula of two specific marginal distributions. The copula kernel operator is essentially a constructor to build a multivariate distribution from a set of univariate distributions. Here, we use the simplest kernel function product. Then, $D(L,\Gamma)=D_L(L) D_\Gamma(\Gamma)$, and, following the steps in Section \ref{amb.sec}, the sought for CDF of eccentricity is 
\eb 
N(L,\Gamma|e<e_0)=16\pi^3\,\int D(L)dL\,\int_{\Gamma_0}^L D_\Gamma(\Gamma) \Gamma d\Gamma.
\label{D.eq}
\ee 
Noting that $\Gamma=L\,\beta$, where $\beta\equiv \sqrt{1-e^2}$, and the number of $\Gamma$ above $\Gamma_0$ at a fixed $L$ is the same as the number of $\beta$ above $\beta_0=\sqrt{1-e_0^2}$, we obtain
\eb 
{\rm CDF}[e]\propto \int_{\beta_0}^1 D_\beta(\beta)\beta \,d\beta.
\ee 
The thermal CDF$(e)$ in Eq. \ref{therm.eq} is obtained under the assumption that $D_\beta(\beta)=1$. This implies that the density of $\beta$ at any fixed $L$ is uniform. A rich variety of marginal eccentricity distribution models can now be formalized by assuming non-uniform  density profiles $D_\beta(\beta)$. For example, a power-law density profile $D_\beta(\beta)=\beta^{-\alpha}$ corresponds to a CDF$(e)= 1-(1-e^2)^{1-\alpha/2}$. The Jeans--Ambartsumian's assumption is a particular case when $\alpha=0$. Positive powers $\alpha$ generate ``superthermal" distributions of eccentricity with concave PDFs rising faster toward $e=1$, while a negative $\alpha$ generates a convex PDF with a maximum at a medium eccentricity.

\begin{figure*}
    \includegraphics[width=0.47 \textwidth]{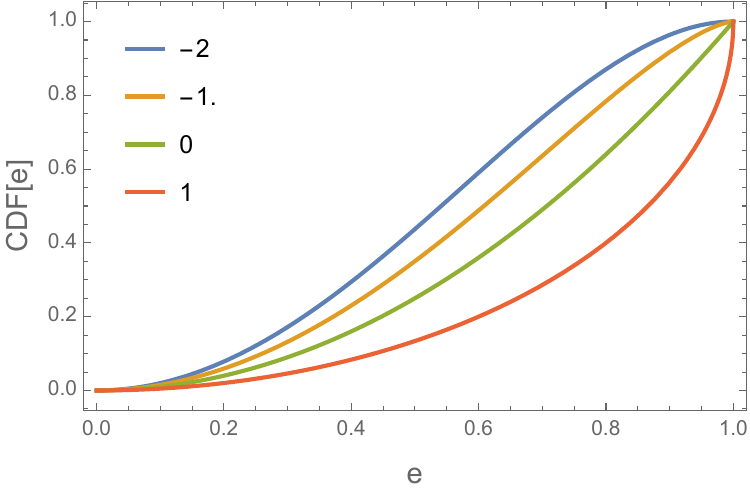}
    \includegraphics[width=0.47 \textwidth]{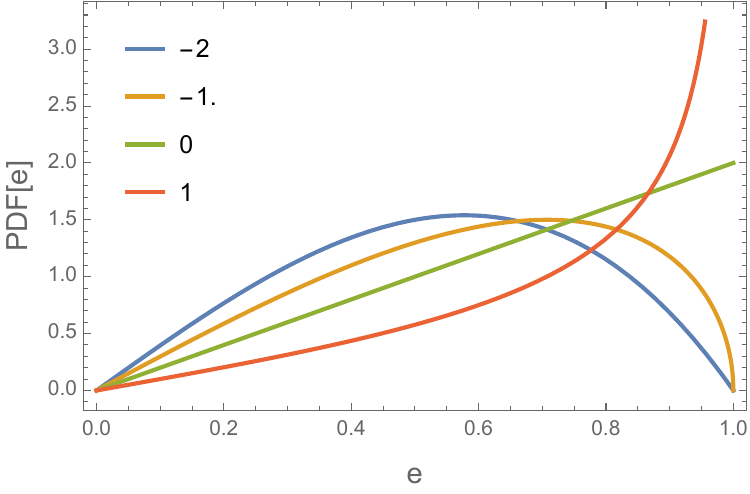}
    \caption{CDF (left panel) and PDF (right panel) of power-law marginal eccentricity distribution for different values of $\alpha'$ per Eq. \ref{pl.eq}.}
    \label{pl.fig}
\end{figure*}

When comparing these models with the observable sample distributions, it is practical to rewrite this generalized approach in the phase space of explicitly independent parameters $\{ {\cal E},\beta,I,M,\omega,\Omega\}$, thus replacing the interrelated Delaunay elements $L$, $\Gamma$, and $H$. The total orbital energy ${\cal E}$ is defined as
\eb 
{\cal E}\equiv -\frac{G m_1 m_2}{2\,a},
\label{E.eq}
\ee 
while $\beta\equiv \sqrt{1-e^2}$ and $I\equiv \cos i$. The basic assumption again is that the phase density $D({\cal E},\beta)$ is a a copula distribution $D_{\cal E}({\cal E})\cdot D'_\beta(\beta)$, where we use $D'$ for the phase density of $\beta$ at a fixed ${\cal E}$ to distinguish it from previously considered $D$ at a fixed $L$.
This readily furnishes the relation
\eb 
{\rm CDF}[e_0]\propto N({\cal E},\beta|\beta>\beta_0) \propto \int_{-\infty}^0 D_{\cal E}({\cal E}) d{\cal E} \int_{\beta_0}^1 D'_\beta(\beta) \,d\beta \propto \int_{\beta_0}^1 D'_\beta(\beta) \,d\beta.
\label{eps.eq}
\ee 
The family of power-law eccentricity distributions is obtained if $D'_\beta(\beta)\propto \beta^{1-\alpha'}$ for $\alpha'<2$:
\begin{eqnarray}
    {\rm CDF}[e]&=&1-(1-e^2)^{1-\alpha'/2}\nonumber \\
    {\rm PDF}[e]&=&(2-\alpha')e(1-e^2)^{-\alpha'/2}.
    \label{pl.eq}
\end{eqnarray}
Fig. \ref{pl.fig} shows the corresponding functions for four different values of $\alpha'$. This model incorporates a wide variety of distributions including bell-shaped and steeply rising PDFs. Again, we note that the Jeans-Ambartsumian's assumption is a particular case with $\alpha'=0$ (green curves).

\section{The Gaia-based sample of resolved binary systems}
\label{gaia.sec}
Based on the data from the third Gaia mission release \citep{2016A&A...595A...1G, 2023A&A...674A...1G}, \citet{2021MNRAS.506.2269E} selected a large, all-sky collection of candidate resolved binary systems (1.3 million pairs). The components of these candidate systems have measured individual positions, proper motions, and parallaxes. The general principles involved in this compilation include consistent parallaxes and proper motions, although the latter criterion should be relaxed due to the expected relative orbital motion. A great deal of effort was directed toward estimation of the remaining contaminants due to chance alignments of field stars and members of stellar clusters and associations. The presence of such interlopers is evident from the sample distribution of the apparent separation
\eb 
s=\rho/\varpi,
\label{rho.eq}
\ee 
where $\rho$ is the angular separation between the components in mas, and $\varpi$ is the measured parallax in mas, cf. Fig. 2 in \citep{2021MNRAS.506.2269E}. Stochastic interlopers dominate the initial sample at $s>30$ KAU, which is seen as a general rise of the logarithmic histogram. The authors proposed a chance alignment probability parameter ${\cal R}$ for each pair, which helps to filter out the contaminants. 
Applying a filter ${\cal R}<0.1$ apparently removes most of the contaminants. For inference on the intrinsic distribution of the widest binaries, it is important to minimize the risk of including chance alignments, and I imposed a much stricter filter ${\cal R}<0.01$. Furthermore, at the cost of removing a major fraction of the initial sample, I found it necessary to impose additional restrictive criteria. The rate of chance alignments is greatly higher for distant stars, where the measured parallaxes are affected by unresolved binarity, for example \citep{2021A&A...649A...5F}. Closely separated double stars in both resolved and unresolved pairs perturb Gaia astrometry making the resulting proper motions unreliable \citep{2017ApJ...840L...1M}. Some widely separated pairs show abnormally high relative velocities, and are found to be associated with elevated values of the {\tt ruwe} parameter, which quantifies the degree of compliance of each source with the 5- or 6-parameter astrometric model. Widely separated companions are often parts of hierarchical triple or multiple systems \citep[e.g.,][]{2004RMxAC..21....7T, 2009ApJ...703.1760M}, where the orbital motion of tight inner pairs generate false proper motion signals. These risks are mitigated by additional filters: {\tt ruwe}$<1.3$, $\rho>2\arcsec$, $\varpi>4$ mas. The remaining sample counts 103,169 pairs.

\section{On the distribution of orbital energy}
\label{ene.sec}
\citet{1919MNRAS..79..408J}, using the initial hypothesis of Boltzmann's distribution of orbital energy in a dynamically relaxed binary population, derived a functional law for the expected marginal distribution of energy, which turned out to be non-physical. This feature was used to refute the starting supposition of thermalisation of binary systems. \citet{1937AZh....14..207A} showed, however, that the ``thermal" distribution of orbital eccentricity (Eq. \ref{the.eq}) can be obtained for any distribution of energy as long as the phase density is dependent only on the first Delaunay parameter $L$. In Section \ref{met.sec}, it is further shown that the phase density that is dependent on energy but decoupled from its marginal distribution still furnishes the same marginal PDF of eccentricity. This leaves the problem of energy distribution open. Further progress on the theoretical side could only be achieved in greatly more sophisticated numerical simulations, which involve mapping the immense parameter space of interacting binaries.

\begin{figure*}
    \includegraphics[width=0.47 \textwidth]{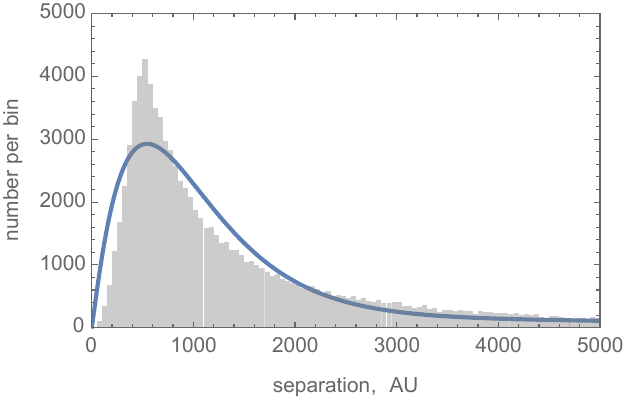}
    \caption{Empirical distribution (histogram) of the projected separations between the components of resolved binary systems in Gaia DR3 from the catalog by \citet{2021MNRAS.506.2269E} after additional filtering. The blue line represents the properly scaled PDF of the best-fitting analytical distribution, which is a $\{0.78,0.22\}$ mixture distribution of Gamma$[2.1,487]$ and LogNormal$[8.68,0.83]$ distributions. The peculiar sample PDF is not adequately represented by any available analytical forms, having a concave ascending slope, a sharp peak, and a exponentially descending heavy tail.}
    \label{sep.fig}
\end{figure*}

\citet{1975MNRAS.173..729H} in his seminal paper suggested a power-law fit to the CDF of the standardized energy. We note that his Eq. 2.20 provides the survival function ($1-$CDF) rather than the CDF. With this caveat in mind, we derive the corresponding Heggie's PDF of orbital energy as
\eb 
D_{\cal E}({\cal E})\propto |{\cal E}|^{-\frac{5}{2}}.
\label{calE.eq}
\ee 
This functional form implies that the vast majority of binaries in a sufficiently dense environment of a star cluster should have energies close to zero, i.e., semimajor axes piling up at the highest possible values. This obviously is not seen in our data (cf. Fig \ref{sep.fig}). The histogram of projected separations (which cannot drastically differ from the distribution of $a$) is sharply peaked at $\sim 550$ AU, and the frequency of larger separations is slowly declining with separations beyond 2000 AU. We should take into account here that soft binaries can be efficiently formed in dense clusters via three-body interactions \citep{1971SvA....15..411A, 2024ApJ...970..112A}, but they can hardly be tightened by subsequent encounters with other stars in the cluster. According to the models developed by Heggie, such encounters between binaries and singles at sufficiently low relative velocity result in further softening of the binaries or their disruption, while the initially hard binaries typically become harder, see also \citep{1975AJ.....80..809H}. A significant external potential can alter the statistical outcome in dense environments \citep{2021MNRAS.508..190G} and even affect the rate of dissipation of such systems---however, if the binaries we see today in the field are the survivors of dissolved clusters, a degree of bifurcation should be seen in the distribution of separations and binding energies. Old binaries in the field should also survive the dynamical interaction in the Galaxy. Chance flybys of field singles can further increase the separation and eccentricity, or completely disrupt a soft binary. The rate of these events depends on the pecuiliar motion and Galactic orbit of the binary. Extremely wide binaries with separations in the range $10^4$--$10^5$ AU can be dissolved by the Galactic tidal potential beyond the Galactic Hill radius \citep{2010MNRAS.401..977J}, although the actual critical distance depends on the orbital rotation alignment, and a counter process of chaotic capture of unrelated stars to form a new binary is physically possible \citep{2012MNRAS.421L..11M}.

The available Gaia collection of soft binaries does not support the theoretical power law in Eq. \ref{calE.eq}, nor is it consistent with the dynamical bifurcation model. This becomes obvious from the observed sample distributions of angular separations in the largest collection of wide binaries available today.

\section{Orbital distributions and observational data}
\label{obs.sec}

Orbits of wide stellar binaries are not known from observation, because their orbital periods are much longer than astronomers' lifetimes. We can only rely on secondary observational data, which are indirectly related to the orbital elements. These come from the astrometric catalogs, with a recent burst of interest in this area of research caused by the unprecedented scope and precision of the Gaia mission products \citep{2016A&A...595A...1G, 2023A&A...674A...1G}.

\citet{2020MNRAS.496..987T} used the observed distributions of the relative proper motion magnitude (converted to relative velocity using the observed parallax for each object) and the direction of this tangential vector to estimate the intrinsic sample distribution of eccentricity in an earlier collection of Gaia measurements. His conclusion was that there is a dichotomy of eccentricity probability density functions (PDF) at different parts of the apparent separation range. While the overall distribution seems to approximately follow the thermal PDF (Eq. \ref{2e.eq}), the harder binaries at separations $<200$ AU have generally smaller eccentricities, while the softer systems at separations $>1000$ AU appear to pile up at $e=1$, and their PDF can be approximated as $e^{1.2}$.

The observable distributions include the projected angular separation $\rho$, which is converted to an estimated projected physical separation in KAU using Eq. \ref{rho.eq}. This parameter is mostly related to the intrinsic distribution of semimajor axes $a$ and thus, to the orbital energy. Unfortunately, the projected separation is also dependent on the PDF of eccentricity $e$. To separate these unknown PDFs, additional measurements are needed. \citet{2004RMxAC..21....7T, 2005AJ....129.2420M} proposed to use the relative projected velocity of the components,
\eb 
\Delta v=4.74\,||\Delta\boldsymbol{\mu}||/\varpi,
\label{dv.eq}
\ee 
where $\Delta\boldsymbol{\mu}$ is the measured proper motion difference between the components. Clearly, this parameter is strongly affected for distant systems, where the combined effect of measurement errors in proper motions and parallax will produce unreasonably high velocity estimates in the statistical sense.

The third observable parameter, which is derived from the astrometric measurements, is the angle $\vartheta$ between the separation vector and the vector of relative velocity, projected on the sky plane. It is called the motion angle in this paper, also known as the v--r angle in the literature \citep{1998AstL...24..178T}. This value can be computed via the relation
\eb 
\cos{\vartheta} = \boldsymbol{v}_t \cdot \boldsymbol{\rho}/(||\boldsymbol{v}_t||\;||\boldsymbol{\rho}||).
\label{cost.eq}
\ee 
The norm of $\boldsymbol{v}_t$ equals to $\Delta v$ in Eq. \ref{dv.eq}. It is practical to separate $\Delta v$ and $\cos{\vartheta}$, although they come from the same observable vector $\boldsymbol{v}_t$, because they map to different sets of intrinsic (physical) distributions. Table \ref{map.tab} displays the mapping relations between the three intrinsic parameters (semimajor axis, eccentricity, and mass) and the three observables (projected separation, relative velocity, and the motion angle). The nonlinear orbital equations involve additional geometric values: inclination $i$, position angle of the ascending node $\Omega$, and the periastron argument $\omega$, if we employ the traditional schema based on the 3-1-3 Euler rotation. However, these distributions are fixed if the orientation of binary orbits is assumed to be random and isotropic. Indeed, the angles $\Omega$ and $\omega$ are then uniformly distributed over $[0, 2\pi]$ interval, while the inclination is distributed as
\eb 
D_i(i)=\frac{1}{2} \sin{i}
\label{sini.eq}
\ee 
over $[0,\pi]$.

Table \ref{map.tab} shows that $\vartheta$ is the ``cleanest" of the observable parameters, in the sense that it depends on a single intrinsic distribution of eccentricity. Within the proposed power-law model and the copula model, this seems to provide the opportunity to estimate the power index $\alpha'$ by direct Monte Carlo simulations. There are significant caveats for this method. High-eccentricity orbits dominate for thermal and super-thermal ($\alpha'\ge 0$) distributions. The majority of pairs should be observed closer to their apocenters, where the separation is larger than $a$, and the relative velocity is slow \citep{2005AJ....129.2420M}. Many systems are also seen nearly edge-on (cf. Eq. \ref{sini.eq}). The combination of these conditions should result in a pile-up of $\vartheta$ at 0 or $\pi$. However, the random measurement error, which is uncorrelated with the separation vector, tends to level up the observed distribution of $\vartheta$.

\begin{figure*}
    \includegraphics[width=0.78 \textwidth]{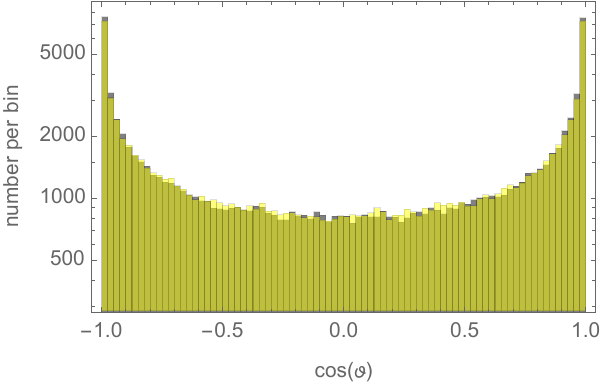}
    \caption{Observed (gray color) and Monte Carlo simulated (yellow color) distributions of the
    cosine motion angle $\cos{\vartheta}$, Eq. \ref{cost.eq}.}
    \label{ugol.fig}
\end{figure*}

Fig. \ref{ugol.fig} shows the ``observed" values $\cos{\vartheta}$ (Eq. \ref{cost.eq}) as a gray-colored histogram and a Monte Carlo simulated set of the same size as a yellow-colored histogram, which assumes a flat (isotropic) distribution of motion angles $\vartheta$. Apparently, the sample distributions are quite close, and the null hypothesis that the motion angles are completely random can be accepted. Unfortunately, the cleanest observable motion angle is also the most vulnerable to random measurement error, as will be discussed in the next section. To some extent, randomization of the motion angle by measurement error is expected to be responsible for the apparent flattness of its distribution.

\begin{table}[h]
\caption{Mapping of observable parameters (first column) and intrisic orbital parameters (upper row) for an ensemble of wide binary stars}
\tablewidth{0pt}
 \hrule
 \begin{tabular}{l|c|c|c|} \label{map.tab}
 & semimajor $a$ & 
eccentricity $e$ &total mass $m$ \\
separation $s$ & \ding{58} & \ding{58} & --\\
velocity $v_t$ & \ding{58} & \ding{58} & \ding{58}\\
motion angle $\vartheta$ & -- & \ding{58} & -- 
\end{tabular}
\hrule
\end{table}

\section{Estimating eccentricity from motion angle}

In principle, the formal uncertainty of the motion angle $\vartheta$ can be computed for each pair of stars in the following way. Let $\boldsymbol{C}_\mu$ be the combined variance-covariance matrix of the proper motion difference $\Delta\boldsymbol{\mu}=[\Delta\mu_{\alpha*},\Delta\mu_\delta]^T$, which is directly computed from the formal errors and correlation coefficients given in the catalog (assuming that the two proper motions are statistically independent). The error-normalized proper motion magnitude in the direction of observed proper motion is then computed as $\sqrt{\Delta\boldsymbol{\mu}^T\, \boldsymbol{C}_\mu^{-1} \, \Delta\boldsymbol{\mu}}$, and the orthogonal component (signal-to-noise ratio) in the perpendicular direction is $Y=\sqrt{\Delta\boldsymbol{\mu}_\tau^T\, \boldsymbol{C}_\mu^{-1} \, \Delta\boldsymbol{\mu}_\tau}$, where $\Delta\boldsymbol{\mu}_\tau=[\Delta\mu_\delta,-\Delta\mu_{\alpha*}]^T$. The reciprocal of $Y$ is approximately the formal uncertainty of the position angle of the proper motion difference vector in radians. Direct computations reveal that the general sample of Gaia pairs has a broad distribution of PA uncertainties peaking at approximately 0.2 rad. The elevated uncertainty of the proper motion directions is related to the prevalence of distant and faint stars with small apparent motion on the sky. This justifies the stringent cuts implemented in the working sample (Section \ref{gaia.sec}). 

In a general, all-inclusive adjustment of intrinsic orbital distributions, we would like to retain the distant pairs, because they are more often associated with the wide separations. In this paper, the formal uncertainty of $\Delta\boldsymbol{\mu}$ is woven in the Monte Carlo simulation procedure in the most direct and consistent way. A random set of scaled geometric configurations is generated starting with a uniform distribution of $\Omega$, $\omega$, and the instantaneous mean anomaly $M$, and a realization of inclination $i$ distributed by Eq. \ref{sini.eq}. The orbital eccentricity is a random number from PDF[$e$] in Eq. \ref{pl.eq} with a fixed power index $\alpha'$. Eccentric anomalies are generated from the uniformly distributed mean anomaly and the synthetic eccentricity values using Kepler's equation\footnote{\url{https://mathworld.wolfram.com/KeplersEquation.html}} and the fast reversed interpolation schema \citep{2018arXiv181202273T}. These input parameters are sufficient to compute a realization of the four relevant Thiele-Innes parameters $A_f$, $F_f$, $B_f$, and $G_f$ with an arbitrary scale $a_0$. From these, the normalized instantaneous value of orbital proper motion $\Delta\boldsymbol{\mu}$ is computed along with the instantaneous apparent separation vector $\boldsymbol{\rho}$. The angle between these vectors provides a random realization of $\vartheta$.

Generating realistic observational error components, which are not scale-invariant, requires transformations of proper motions to velocities and a prior on the distribution of orbital separations. The measurement noise is taken into account by random number generation of vectors $\boldsymbol{\nu}$ from the bivariate normal distribution with the specific (randomly chosen) covariances $\boldsymbol{C}_\mu$, which are added to the synthetic vectors $\boldsymbol{v}_t$ with appropriate scaling for distance and unit conversion from AU yr$^{-1}$ to km s$^{-1}$. The parallaxes needed for this operation are not synthetic but are taken from the corresponding data. The same specific parallaxes are also used to convert the normalized synthetic proper motion vectors to tangential velocities, assuming the empirical distribution of orbit sizes shown in Fig. \ref{sep.fig}. The 103,169 random relative velocity vectors $\boldsymbol{v}_t$ with added measurement noise and projected separation vectors $\boldsymbol{\rho}$ are used to compute the synthetic motion angles and compare with the same number of available data points. The net result is the sample distribution of motion angles and the concordance value L1F. This cycle is completed for a grid of eccentricity indices $\alpha'$. This procedure is repeated 109 times for a uniform grid of $\alpha'$ values from -1.50 to +0.2. 

The next task is to evaluate how well the generated set of $\vartheta$ matches the observed data set. The empirical distribution of $\vartheta$ is quite flat and well approximated by a rectangular function. Any departures from uniformity in the synthetic set would indicate a model inconsistency. Two sample distributions are often compared using the non-parametric Kolmogorov-Smirnov test. It belongs to the class of rank-order statistical methods, where the two CDFs, which can be both empirical, are matched by the supremum of their absolute difference. While based on elegant mathematical theory, this method may have a diminished practical value in some special cases, when the empirical distribution has a confined feature such as a pile-up at a marginal value (which is the case with the empirical distribution of separations $s$). The supremum estimator is not robust in the presence of distribution features. A different, more robust and stable test criterion called L1F is proposed here. If $H_1(X)$ and $H_2(X)$ are the two CDFs of a random variate $X$ to be compared, then
\eb 
{\rm L1F}=\int_X |H_1(X)-H_2(X)|dX,
\ee 
where the integral is taken over the entire range of $X$. Thus, the supremum of the Kolmogorov-Smirnov test is replaced by the mean absolute difference. Technically, since the empirical distribution is discrete, the computation involves sorting both data sets, finding for each element of one sorted sequence the nearest element in the other sorted sequence, summing up the absolute differences of the indices of the matched couples, and dividing the result by the sample size. The L1F value is limited by zero on the low end, but the maximum possible value is equal to the sample size, if both sets are of equal size. The method can be trivially formalized for samples of unequal size or continuous test distributions by appropriate normalization of the indices. The match is ideal when L1F$=0$.

\begin{figure*}
    \includegraphics[width=0.58 \textwidth]{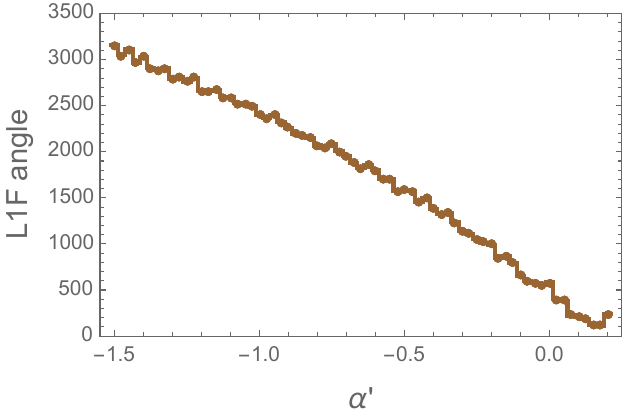}
    \caption{Distribution concordance metric L1F of Monte Carlo random samples of orbital configurations with the empirical sample from Gaia GR3 data in the motion angle parameter $\vartheta$ versus the assumed power index $\alpha'$ of the intrinsic eccentricity PDF. Small L1F values indicate a nearly perfect match of the distributions.}
    \label{l1f.fig}
\end{figure*}

The result of this Monte Carlo adjustment is shown in Fig. \ref{l1f.fig}. It depicts the evolution of the L1F metric for the closeness of the empirical distribution of motion angles $\vartheta$ and the randomly generated synthetic samples of orbit configurations on a grid of $\alpha'$. The large L1F values at $\alpha'<0$ suggest that the corresponding convex PDFs of eccentricity within the copula phase density model do not provide a good match of the observed and synthetic motion angles. Specifically, the observed distribution of $\vartheta$ is quite flat and concave, with a barely noticeable depression at the center, whereas the distribution simulated with a positive $\alpha'$ is distinctly convex with a broad bump at $90\degr$. In other words, if the soft binaries from Gaia were mostly low-eccentricity systems, we would see a preponderance of nearly orthogonal motion angles between separation and relative proper motion vectors, which is not the case.
The best-fitting value of $\alpha'$ is +0.15, which is a purely empirical result not related to any theoretical scenarios reviewed in Section \ref{int.sec}.

\section{Conclusions}
\label{end.sec}

The origin of steeply rising distribution functions of eccentricity for wide binaries in the field has long been a discussion topic in the astronomical literature, because it provides a powerful method to validate basic theoretical models and identify possible tensions in our star formation and dynamical evolution models. The early heuristic derivation of the so-called thermal probability density from  first-principle assumptions about orbital energy equipartition due to interactions between the stellar systems proved to be both theoretically problematic and rather consistent with observations. The intrinsic inconsistency of the proposed model was noted by its author \citep{1919MNRAS..79..408J}, because it required a rapidly declining PDF of the total orbital energy, i.e., a preponderance of marginally wide binaries. \citet{1937AZh....14..207A} noted that the same linearly rising PDF of eccentricity obtains with a more general and accommodating assumption that the phase density of orbital configurations depends only on the first Delaunay element $L$ (implicitly, on the masses and semimajor axes) in an unspecified way, not requiring a specific distribution law for the total energy. In this paper, we have shown that this approach can be further generalized for the case of a copula phase density distribution, which combines two arbitrary marginal distributions of orbital eccentricity and energy via a product kernel function. This opens the possibility to fit a certain integrable mathematical model (i.e., a functional form) of the intrinsic PDF$[e]$ and compare the resulting multi-dimensional sample distribution with observations without any assumptions about the distribution of energy. A power-law distribution model per Eqs. \ref{pl.eq} is practical and versatile, because it incorporates a range of convex and concave distribution laws specified by a single power index $\alpha'$, and the much discussed thermal PDF$[e]$ is a particular case within this model.

\citet{2022ApJ...929L..29H} considered the possible evolution scenarios of wide binaries in the Galactic field driven by tidal perturbations. Assuming that the orbits are isotropic in the Galactic coordinate system, a superthermal distribution of eccentricity ($\alpha'>0$) can only be the result of a superthermal initial distribution. This presents a conundrum to be solved: if most of the binaries are formed within relatively dense clusters, the encounters between the cluster members should predominantly disrupt the widest and most eccentric systems \citep{2023ApJ...955..134R}. After the cluster dissolution, the surviving systems should preferably be harder and less eccentric. Yet, we find a finite number of extremely wide binaries in the field with separations stretching to $\sim1$ pc. The accuracy of the Gaia mission products has reached the levels sufficient to probe the orbital dynamics parameters for large samples including $10^5$--$10^6$ systems. This can be done by a direct simulation of the observable values (projected separation, projected relative velocity, and the motion angle) from assumed distributions of the intrinsic parameters using the Monte Carlo method. The motion angle stands out as the ``easiest" parameter, because it maps to only one truly unknown distribution, which is the PDF of eccentricity (Table 1). Thus, without any explicit assumptions about the distribution of orbit sizes and masses, we can simulate a large sample of synthetic pairs for a fixed $\alpha'$ and compare the emergent sample distribution of motion angle $\vartheta$ (or its cosine) with the Gaia DR3 dataset. Repeating this analysis for a grid of $\alpha'$ and minimizing the CDF difference via, for example, the L1F metric per Eq. \ref{l1f.fig}, we determine the best-matching power law of eccentricity.

The numerical experiments described in this paper have revealed that the available sample of motion angles for $10^5$ carefully filtered Gaia-resolved binaries is almost perfectly consistent with a superthermal PDF$[e]$ at $\alpha'=+0.15$ (Fig. \ref{l1f.fig}). This follows from the nearly flat sample distribution of the measured motion angles. This conclusion is in accord with previous analyses for smaller samples of resolved binaries \citep{2016MNRAS.456.2070T, 2022MNRAS.512.3383H}, but not to the extreme option suggested by \citet{2022ApJ...933L..32H}, where a delta-function PDF at unity is considered. 
This results comes with a caveat, however. A significant fraction of the wide resolved systems in Gaia is expected to include hierarchical multiples. The unresolved inner pairs with much shorter orbital periods may significantly perturb the observed proper motions of photocenters leading to abnormally large relative velocities. This proves to be the crucial difficulty for unambiguously testing the alternative theories of gravitation like MOND on the Gaia sample of wide binaries \citep{2023OJAp....6E...2M, 2023OJAp....6E...4P}. Unfortunately, the motion angle method is also vulnerable to this effect. The additional velocity component from an unresolved inner companion randomizes the angle and contributes to the observed isotropic probability density (Fig. \ref{ugol.fig}). The applied strict filter {\tt ruwe} $<1.3$ is expected to clean the data set of many such contaminants, but the remaining fraction of multiples can still bias the result. For a dataset of this size ($10^5$ pairs), only statistical methods of mitigating this bias are practical, but they require scrupulous modeling of many involved physical parameters and a general knowledge of the contamination frequency. In addition to the physical dispersion of relative proper motion vectors, excessive measurement noise can flatten the observed distribution too. Special numerical experiments with all formal errors of proper motion components scaled by common factors 1.1 and 1.2, while keeping the RA-Dec correlation coefficients as given, revealed that this method is insensitive to the underestimated formal errors, still providing best-fitting power index parameters $\alpha'$ above zero.

The generalized statistical approach and the fits of power-law eccentricity models using the observed motion angle distribution presented in this paper do not allow us to gain new information on the dynamical masses in the wide binary systems. The component masses are explicitly tangled with the orbit sizes in both the Delaunay $L$ and orbital energy ${\cal E}$ in Eqs. \ref{D.eq} and \ref{eps.eq}. The marginal sample distribution of $a$ can be independently estimated by solving an ill-posed deconvolution problem with appropriate regularization, for a fixed eccentricity power-law model. According to Table 1, the only remaining observable parameter suitable for separating the mass (or energy) distribution is the relative velocity $v_t$. However, given the apparent dominance of highly eccentric orbits, the accuracy of this method may be limited by the low relative motions of systems in the protracted apoapsis phases.

\section*{Acknowledgements}
I would like to thank the anonymous referee for constructive comments and suggestions that significantly improved the clarity of presentation in this manuscript.

\bibliography{main}
\bibliographystyle{aasjournal}

\end{document}